\documentclass[12pt]{JHEP3}

\usepackage{amsmath}
\usepackage{longtable}
\usepackage{caption}

\newcommand{\be}{\begin{eqnarray}}
\newcommand{\ee}{\end{eqnarray}}

\newcommand{\bn}{\begin{enumerate}}
\newcommand{\en}{\end{enumerate}}

\def\Tr{{\rm Tr}}

\def\det{{\rm det}}

\newcommand{\ZZ}{{\mathbb Z}}

\newcommand{\RR}{{\mathbb R}}

\newcommand{\cN}{{\mathcal N}}
\newcommand{\sign}{{\rm sign}}
\newcommand{\tX}{{\tilde X}}
\newcommand{\ra}{\rightarrow}
\newcommand{\Hom}{{\rm Hom}}
\newcommand{\cW}{{\mathcal W}}

\title{Dualities for 3d Theories with Tensor Matter}

\author{ Anton Kapustin $^{1}$, Hyungchul Kim $^{2}$,  Jaemo Park$^{2,3}$

\\
$^1$ California Institute of Technology, Pasadena, CA 91125, USA
\\

$^2$Department of Physics, POSTECH, Pohang 790-784, Korea
\\
$^3$Postech Center for Theoretical Physics (PCTP), Postech, Pohang
  790-784, Korea

\\
\\
E-mail: \email{kapustin@theory.caltech.edu, dakiro@postech.ac.kr,
jaemo@postech.ac.kr} } 

\abstract{ We study dualities for ${\cal N}=2$ 3d Chern-Simons matter
theories with gauge groups U/Sp/O, matter in the two-index tensor representations
(adjoint/symmetric/antisymmetric) in addition to the fundamental representation, and a superpotential. These dualities
are analogous to Kutasov-Schwimmer-Seiberg dualities in 4d. We test them by computing the superconformal index and
the partition function on $S^3$ for many dual pairs and find
perfect agreement. In some cases we find a simple dual description for theories with tensor matter and
no superpotential, thereby generalizing the "Duality Appetizer" of Jafferis and Yin to an infinite class
of theories. We also investigate nonperturbative truncation of the chiral ring proposed in the context of 4d dualities.}

\begin{document}

\section{Introduction}

Recently, there has been renewed interest in nonperturbative dualities between
three dimensional theories such as mirror symmetry and Seiberg-like
dualities. This is explained in part by the availability of
sophisticated tools such as the partition function on $S^3$ and the
superconformal index. Using
these tools, one can give impressive evidence
for various 3d dualities. Some of works in this area are \cite{Giveon09}-\cite{Aharony11}. One can also obtain
the R-charge of the fields by maximizing the free energy of the
theory of interest \cite{Jafferis10}.

In this paper we continue this line of research and study Seiberg-like dualities
for $\cN=2$ $d=3$ gauge theories with classical gauge groups and matter both in the fundamental and the two index tensor
representations. Many of these proposed dualities can be motivated using the Hanany-Witten brane
setup and brane moves passing through configurations with coincident NS5 branes \cite{Elitzur97,GiveonKutasov98}. Similar dualities for $\cN=1$ $d=4$ theories have been studied in the 90's by Kutasov and collaborators \cite{Kutasov95,KutasovSchwimmer95,KutasovSchwimmerSeiberg95} and others \cite{Intriligator95,LeighStrassler95,IntriligatorLeighStrassler95}.
We explore various 3d gauge theories with classical groups U/O/Sp and
with matters in adjoint/symmetric/antisymmetric representations combined with
fundamental representations and propose dual descriptions for them. We
subject these dualities  to various tests such as the comparison of the $S^3$ partition
functions and the superconformal indices.

Theories with matter in the tensor representations are far more complicated than those with just the fundamental matter. In 4d, to simplify the analysis, the superpotential for the matter in tensor representations is introduced so that the
chiral ring is truncated. To this date, the dual description of such theories without the superpotential is not known. In 3d one can also introduce the Chern-Simons coupling. In some cases we propose a dual description of a 3d gauge theory which has no superpotential but has a nonzero Chern-Simons coupling instead. The simplest example of this sort is essentially the "Duality Appetizer" found by Jafferis and Yin \cite{JafferisYin11}. We propose several infinite series of dualities which generalize the "Duality Appetizer". In all these dualities a nontrivial looking
nonabelian gauge theory is mapped to a theory of free chiral multiplets. From the viewpoint
of the nonabelian gauge theory, the R-symmetry of the free chiral multiplets
is {\it accidental} (the Jafferis-Yin example is an exception in this regard). This gives a plethora of models where the F-maximization
procedure \cite{Jafferis10, Klebanov11}  fails because of the presence of an accidental R-symmetry.

Another peculiar feature worthy of mention is the nonperturbative truncation
of the chiral rings. It has been proposed in \cite{KutasovSchwimmer95} that quantum effects lead to new relations in the chiral ring. Without this quantum truncation the duality cannot work. The index computation allows us to see that such truncation does indeed occur in 3d $\cN=2$ theories and clarifies its mechanism.

The content of the paper is as follows. In section 2, we briefly review the tools we use (the partition function on $S^3$ and the
superconformal index). In section 3, we handle seven cases of
Seiberg-like dualities for theories with matter in both tensor and fundamental
representations and a superpotential.  All these dualities have 4d analogues, cf.  \cite{IntriligatorLeighStrassler95}. We give evidence for the conjectured dualities by working out the partition function and the superconformal index. For three classes of these theories we also propose a dual description of the theory without the superpotential thereby generalizing the Jafferis-Yin duality \cite{JafferisYin11}.
Section 4 contains some additional remarks.

\section{The partition function and the superconformal index}

The $S^3$ partition function for $\cN=2$ $d=3$ superconformal gauge theories with general R-charges was worked out in \cite{Jafferis10,HamaHosomichiLee10} by generalizing the localization approach of \cite{KWY1}. Consider a theory with
gauge group $G$ and chiral multiplets in representations $R_i$. Such a theory typically flows to an IR fixed point at which chiral multiplets can have
noncanonical conformal dimensions. $\cN=2$ superconformal symmetry requires the dimension of a chiral field to be equal to its R-charge, so instead of conformal dimensions we can equally well talk about R-charges. Let us denote the R-charge
of the scalar field in the chiral multiplet by $\Delta_i$. The
partition function on $S^3$ is given by
\begin{equation}
Z=\int \prod_{Cartan}  e^{i\pi \Tr\, u^2} \prod_{\rho(u)\in R_G}(2\sinh
\pi \rho(u)) \prod_{\rho(u)\in  R_i}
e^{l(1-\Delta_i+i\rho(u))}.
\end{equation}
The integration is over the Cartan subalgebra of the gauge group $G$. For example,
for $G=U(N)$ we have integration variables $u_a$ with $a=1,\ldots,
N$. Tr denotes the trace over  Chern-Simons term, which is
normalized such that for $U(N)$ at Chern-Simons level $k$, it is $k$
times the ordinary trace.  The factor $\prod_{\rho(u)\in G}(2\sinh \pi \rho(u))$ comes
from the one-loop contribution from the vector multiplets. The product runs
over the roots of the gauge group $G$.
Thus for $G=U(N)$ we get a factor
$\prod_{1\leq a < b\leq N}(2\sinh \pi (u_a-u_b))^2$. The remaining
terms come from the 1-loop contribution of matter multiplets, where the product
runs over all matter representations $R_i$ and all weights of $R_i$.
The function $l(z)$ is given by
\begin{equation}\label{eq:lz}
l(z)=-z\, {\rm log} (1-e^{2\pi iz})+\frac{i}{2} \left(\pi
z^2+\frac{1}{\pi}{\rm Li}_2(e^{2\pi i z})\right)-\frac{i\pi}{12}.
\end{equation}
It can also be defined as the unique solution of $dl(z)/dz=-\pi z \cot(\pi z)$ satisfying $l(0)=0$. Thus for $G=U(N)$ one chiral multiplet
in the fundamental representation contributes a factor $\prod_{a=1}^N
e^{(l(1-\Delta+iu_a))}$. As argued in \cite{Jafferis10},
the partition functions of dual theories should agree as a function of R-charges $\Delta_i$. If a superpotential is present, there is an additional constraint on $\Delta_i$ coming from the requirement that the superpotential have R-charge $2$.

Next let us discuss the superconformal index for  $\cN=2$ $d=3$
superconformal field theories (SCFT). The bosonic subgroup of the
3-d $\cN=2$ superconformal group  is $SO(2,3) \times SO(2) $. There
are three Cartan elements denoted by $\epsilon, j_3$ and $R$ which
come from three factors $SO(2)_\epsilon \times SO(3)_{j_3}\times
SO(2)_R $ in the bosonic subgroup, respectively. The superconformal
index for an $\cN=2$ $d=3$ SCFT is defined as follows
\cite{Bhattacharya09}:
\begin{equation}
I(x,y)=\Tr (-1)^F \exp (-\beta'\{Q, S\}) x^{\epsilon+j_3}\prod_j y_j^{F_j}
\label{def:index}
\end{equation}
where $Q$ is a special  supercharge with quantum numbers $\epsilon =
\frac{1}2, j_3 = -\frac{1}{2}$ and $R=1$, and $S= Q^\dagger$.  The trace is taken over the Hilbert space
in the SCFT on $\mathbb{R}\times S^2$ (or equivalently over the space of local gauge-invariant operators on $\RR^3$). The operators $S$ and $Q$
satisfy the following anti-commutation relation:
\begin{equation}
 \{Q, S\}=\epsilon-R-j_3 : = \Delta.
\end{equation}
As usual, only BPS states satisfying the bound $\Delta =0 $ contribute to
the index, and therefore the index is independent of the parameter $\beta'$. If we have
additional conserved charges $F_j$ commuting with the chosen supercharges
($Q,S$), we can turn on the associated chemical potentials $y_j$, and then the
index counts the algebraic number of BPS states weighted by their quantum numbers.

The superconformal index is exactly calculable using the
localization technique \cite{Kim09,Imamura11}.  It can be written in
the following form:
\begin{multline}\label{index}
I(x,y)=\\
\sum_{m} \int da\, \frac{1}{|\cW_m|}
e^{-S^{(0)}_{CS}(a,m)}e^{ib_0(a,m)} \prod_j y_{j}^{q_{0j}(m)}
x^{\epsilon_0(m)}\exp\left[\sum^\infty_{n=1}\frac{1}{n}f_{tot}(e^{ina},
y^n,x^n)\right]
\end{multline}

The origin of this formula is as follows.
 To compute the trace over the Hilbert space on $S^2\times\RR$, we use path-integral on $S^2\times S^1$ with suitable boundary conditions
on the fields. The path-integral is evaluated using localization, which means that we have to sum or integrate over all BPS saddle points. The saddle points are spherically symmetric configurations on $S^2\times S^1$ which are labeled by magnetic fluxes on $S^2$ and holonomy along $S^1$. The magnetic fluxes are denoted by  $\{m\}$ and take values in the cocharacter lattice of $G$ (i.e. in $\Hom(U(1),T)$, where $T$ is the maximal torus of $G$), while the eigenvalues of the holonomy are denoted $\{ a
\}$ and take values in $T$. $S_{CS}^{(0)}(a,m)$ is the classical action for the
(monopole+holonomy) configuration on $S^2\times S^1$, $\epsilon_0(m)$
is the Casmir energy of the vacuum state on $S^2$ with magnetic flux $m$, $q_{0j}(m)$ is the $F_j$-charge of the vacuum state, and $b_0(a,m)$ represents the contribution coming from the electric charge of the vacuum state. The last factor comes from taking the trace over a Fock space built on a particular vacuum state. $|\cW_m|$ is the order of the Weyl group of the part of $G$ which is left unbroken by the magnetic fluxes $m$ . These ingredients in the formula for the index are given by the following explicit expressions:
\begin{eqnarray}
&&S^{(0)}_{CS}(a,m) = i \sum_{\rho\in R_{F}} k \rho(m) \rho(a) , \\
&&b_0(a,m)=-\frac{1}{2}\sum_\Phi\sum_{\rho\in R_\Phi}|\rho(m)|\rho(a),\nonumber\\
&&q_{0j}(m) = -\frac{1}{2} \sum_\Phi \sum_{\rho\in R_\Phi} |\rho(m)| F_j (\Phi), \nonumber \\
&& \epsilon_0(m) = \frac{1}{2} \sum_\Phi (1-\Delta_\Phi) \sum_{\rho\in
R_\Phi} |\rho(m)|
- \frac{1}{2} \sum_{\alpha \in G} |\alpha(m)|, \nonumber\\
&& f_{tot}(e^{ia},y,x)=f_{vector}(e^{ia},x)+f_{chiral}(e^{ia},y,x),\nonumber\\
&& f_{vector}(e^{ia},x)=-\sum_{\alpha\in G} e^{i\alpha(a)} x^{|\alpha(m)|},\nonumber \\
&& f_{chiral}(e^{ia}, y,x) = \sum_\Phi \sum_{\rho\in R_\Phi}
\left[ e^{i\rho(a)} \prod_j y_{j}^{F_{j}}
\frac{x^{|\rho(m)|+\Delta_\Phi}}{1-x^2}  -  e^{-i\rho(a)}
\prod_j y_{j}^{-F_{j}} \frac{x^{|\rho(m)|+2-\Delta_\Phi}}{1-x^2}\nonumber
\right]\label{universal}
\end{eqnarray}
where $\sum_{\rho\in R_F}, \sum_\Phi$, $\sum_{\rho\in R_\Phi}$ and $\sum_{\alpha\in G}$
represent summations over all fundamental weights of $G$, all chiral multiplets, all weights of the representation $R_\Phi$, and
all roots of $G$, respectively. For $G=O(N)$  we need to carry out an additional $\ZZ_2$ projection
corresponding to an element of $O(N)$ whose determinant is $-1$. This is explained in
\cite{HKPP11}.







\section{3d dualities with tensor matter}

\subsection{$U(N)$ with an adjoint}
The following duality has been proposed in \cite{Niarchos08, Niarchos, Niarchos11}:
\begin{itemize}
\item
Electric theory: $U(N_c)_k$ gauge group, $N_f$ pairs of fundamental/anti-fundamental chiral superfields $Q^a$,
$\tilde{Q}_b$(where $a$, $b$ denote flavor indices), an adjoint superfield
 $X$, and the superpotential $W_e=\Tr\, X^{n+1}$.

\item
Magnetic theory: $U(n(N_f + k) - N_c)_{-k}$  gauge theory, $N_f$ pairs of fundamental/anti-fundamental chiral
superfields $q_a$, $\tilde{q}^a$, $N_f \times N_f$ singlet superfields
 $(M_{j})^{a}_{b}$, $j=1,\cdots,n$,
 an adjoint superfield $Y$ and a superpotential $W_m=\Tr\, Y^{n+1}+\sum_{j=1}^{n} M_j \tilde{q} Y^{n-j} q$.

\end{itemize}
Here it is assumed that $k> 0$. We will call this the Niarchos duality. It is a 3d analog of the Kutasov-Schwimmer-Seiberg duality \cite{Kutasov95,KutasovSchwimmer95,KutasovSchwimmerSeiberg95}. In the absence of a superpotential these theories are superconformal, with scaling dimensions of the fields depending on $k$. It is assumed that $n$ is such that the superpotential is a relevant perturbation of both electric and magnetic theories which drives the theory to a new IR fixed point. Below we will compute in some cases the dimension of $X$ using $F$-maximization and determine the range of $n$ for which this is true.

We denote the R-charge of $Q$ by $R(Q)=r$. Superconformal R-charges
of both $X$ and $Y$ are $\frac{2}{n+1}$ due to the superpotential.
Duality is supposed to map chiral ring generators $\Tr\, X^j$ to
$\Tr\, Y^j$, $j=1,\ldots,n-1$, and $Q^a X^{j-1} \tilde{Q}_b$ to
$(M_{j})^{a}_{b}$, $j=1,\ldots, n$. This determines
$R(q)=\frac{2}{n+1}-r, R(M_j)=\frac{2(j-1)}{n+1}+2r$. Note that for
$n=1$, one can integrate out $X$ and $Y$, and the conjectured
duality reduces to Seiberg-like duality with only fundamental and
anti-fundamental fields \cite{Giveon09}.

Let us first consider $N_f=0$ and $k=1$. The proposed duality says
that $U(N_c)_1$ Chern-Simons theory with an adjoint $X$ and $W=\Tr\,
X^{n+1}$ is dual to $U(n-N_c)_{-1}$ theory with an adjoint $Y$ and
$W=\Tr\, Y^{n+1}$, where the subscript of the gauge group denotes
the Chern-Simons level. An intriguing special case is $n=N_c$, when
the duality conjecture says that the theory with gauge group
$U(N_c)_1$, an adjoint superfield $X$ and $W=\Tr\, X^{N_c+1}$ is
dual to a trivial CFT (i.e. the theory is massive). Also, for
$n<N_c$ the rank of the magnetic group is negative, which we
interpret as a sign of spontaneous breaking of supersymmetry, as in
\cite{ABJ}.

To understand what is going on, let us set $N_c=2$, i.e. let us
consider $U(2)_1$ with an adjoint and a superpotential $W=\Tr\,
X^3$. Since $U(2)=(SU(2)\times U(1))/\ZZ_2$ and $X$ is neutral with
respect to the $U(1)$ subgroup, the $U(1)$ part of the gauge
multiplet is essentially a trivial TQFT and on a flat space-time can
be ignored. Further, $v=\Tr\, X$ is a singlet with respect to the
gauge group and couples to the rest of the theory only through the
superpotential. On the other hand, Jafferis and Yin
\cite{JafferisYin11} provided strong evidence that $SU(2)_1$
theory with an adjoint and no superpotential is dual to a single
free superfield. This superfield is dual to the composite superfield
$u=\Tr\, (X-\frac12 v)^2$.The superpotential deformation $W=\Tr\,
X^3$ can be expressed in terms of $v$ and $u$ as
$W=\frac{1}{4}v^3+\frac{3}{2}uv$. Appealing to the Jafferis-Yin
duality, we see that in flat spacetime $U(2)_1$ theory with $W=\Tr\,
X^3$ is dual to a theory of two free superfields $u,v$ perturbed by
the above superpotential. This superpotential has a unique
nondegenerate critical point $u=v=0$, hence it makes the theory
massive, as predicted by the duality. Furthermore, it is now clear
why deformations by $W=\Tr\, X=v$ and $W=\Tr\, X^2=u+\frac12 v^2$
break supersymmetry: such superpotentials do not have critical
points when written in terms of free fields $u$ and $v$.

Next one may ask if the duality continues to hold on more general manifolds. It was already pointed out
in \cite{JafferisYin11} that the $S^3$ the partition functions of $SU(2)_1$ with an adjoint and a free theory
differ by a factor $1/\sqrt 2$ and that this factor can be interpreted as a partition function of a decoupled
$U(1)$ Chern-Simons theory at level $2$. That is, it was proposed in \cite{JafferisYin11} that perhaps the dual
of $SU(2)_1$ with an adjoint is a free chiral superfield times the $U(1)_2$ TQFT. We propose an alternative interpretation of this factor. We move it to the other side and interpret $\sqrt 2$ as the partition function of $U(1)$ Chern-Simons theory at level $1/2$. This is suggested by the fact that  on $S^3$, $U(2)_1$ theory is equivalent to $SU(2)_1 \times U(1)_{1/2}$ theory.\footnote{On a non-simply-connected manifold, a half-integral level would make the $U(1)$ Chern-Simons action not gauge-invariant, but on $S^3$ it does not cause any problems because $\pi_1(X)=0$, and therefore all gauge transformations are topologically trivial.} We propose that $U(2)_1$ with an adjoint and $W=0$ is dual to a theory of two free superfields $u$ and $v$. As explained above, this leads to the correct behavior when perturbed by the superpotential $W=\Tr\, X^{n+1}$ for $n=0,1,2$ (supersymmetry breaking for $n=0,1$ and massive theory for $n=2$).

A generalization of the Jafferis-Yin duality to $N_c>1$ should now be obvious. We propose that $U(N_c)_1$ gauge theory with an adjoint $X$ and vanishing superpotential is dual to a theory of $N_c$ free fields $u_1,\ldots,u_{N_c}$. The moduli spaces of both theories are $N_c$-dimensional, and we identify $u_i$ with $\Tr\, X^i$. Clearly, for any $i$ perturbing the magnetic side by $W=u_i$ leads to supersymmetry breaking, in agreement with the fact that perturbing the electric theory by $W=\Tr\, X^i$, $i=1,\ldots, N_c,$ leads to supersymmetry breaking. Furthermore, it is easy to see that $W=\Tr\, X^{N_c+1}$ expressed as a function of $u_i$ has a unique nondegenerate critical point $u_1=\ldots=u_{N_c}=0$, which means that perturbing the magnetic theory by such a superpotential leads to a massive theory, again in agreement with the conjectured duality.

The simplest test of the generalized Jafferis-Yin duality is to consider the low-energy limit of the theory
at a generic point in the moduli space. Generically, the gauge group is broken down to $U(1)^{N_c}$, with all the $U(1)$ factors having Chern-Simons level $1$. Since $U(1)$ Chern-Simons theory at level $1$ is essentially trivial \cite{Witten sl2z}, the low-energy effective theory is the theory of $N_c$ free fields (the moduli).\footnote{The proposal of \cite{JafferisYin11} that $SU(2)_1$ theory with an adjoint is dual to a free theory times a decoupled $U(1)_2$ TQFT also passes this test, since along the moduli space $SU(2)_1$ is broken down to $U(1)_2$, times the decoupled modulus field.}

Generalized Jafferis-Yin duality can be further tested by computing the partition function on
$S^3$ or the superconformal index. The following subtlety has to be kept in mind. If we denote the R-charge
of $X$ to be $r$, then the superfields
$\Tr\, X, \Tr\, X^2, \cdots , \Tr\, X^{N_c}$ have R-charge
\begin{equation}
r,2r, \cdots,N_c r  \label{Rcharge}
\end{equation}
respectively. However, we know that free chiral multiplets have
R-charge $\frac{1}{2}$. Thus on the electric side the correct
superconformal R-symmetry is an accidental symmetry which cannot be
seen on the classical level. Consequently when we test the above
conjecture, on the magnetic side we should use the non-canonical
R-charges for chiral multiplets of the free theory as specified in
eq. (\ref{Rcharge}). We performed this computation numerically for
$N_c=3$, the results are displayed in Table \ref{table1}. To the
accuracy achieved, the results agree with the duality prediction.
For higher $N_c$ the $S^3$ partition function is difficult to
evaluate because the integrand oscillates rapidly.

\begin{table}[h]
\begin{tabular}{|c|cccccc|}
\hline
$r $           & 0.2    & 0.3    & 0.4    & 0.5    & 0.6    & 0.7 \\
$\log |Z|$ & -0.613634 & -0.635679 & -0.318126  & 0.000000 & -0.163086 &  -0.326744 \\
\hline
\end{tabular}
\caption{$S^3$ partition function of the $U(3)_1$ theory with an adjoint superfield $X$ and $W=0$ as a function of $r=R(X)$. The result agrees to the accuracy achieved with the partition function of three free chiral superfields with R-charges $r$, $2r$, and $3r$.\label{table1}}
\end{table}

We also worked out the superconformal index for these theories. It is easy to check that the superconformal index of the $U(N_c)_1$ theory with matter in the adjoint is the same as that of the $SU(N_c)_1$ theory with an adjoint and an extra free superfield (i.e. the $U(1)$ factor in the gauge group may be simply ignored). For $N_c=2,3$ we get:

\begin{enumerate}
\item $SU(2)_1$ with one adjoint X of R-charge  $r$ $\thickapprox$ One free chiral multiplet dual to $\Tr\, X^2$ with R-charge $2r$

$I = 1-x^2-2 x^4-2 x^6-2 x^8+x^{4 r}+x^{6 r}+x^{8 r}+x^{2 r}
(1-x^4-2 x^6 )+x^{-2 r}  (-x^2-x^4+x^8 )+x^{-4 r} (x^6+x^8 )+\ldots$

\item $SU(3)_1$ with one adjoint X of R-charge  $r$ $\thickapprox$ Two free chiral multiplets dual to $\Tr\, X^2$,
$\Tr\, X^3$ with
R-charge $2r$, $3r$

$I = 1-2 x^2-3 x^4+x^{4-5 r}-x^{2-3 r}-x^{2-2 r}+x^{4 r}+x^{5 r}+2
x^{6 r}-2 x^{2+r}+x^{3 r}  (1-2 x^2 )+x^{2 r}  (1-x^2 )+x^{-r}
 (-x^2-x^4 )+\ldots$
%

\end{enumerate}

For $N_f>0$ we cannot get rid of the $U(1)$ factor so easily. We test the duality for several low values of $N_f,N_c$ and $k$ by computing the superconformal index, the results are displayed in the Table 2. In all cases we checked the duality is confirmed. For $N_f=0$ we omit some duplicate cases which differ from the one displayed only by a change of sign of $k$.
\begin{longtable}{|c|c|c|p{8cm}|}
\hline
$(N_f,k,N_c)$       & Electric    &   Magnetic                    &  \\
$n=2,  N_f+k=1$ & $U(N_c)$    &   $U(2 - N_c)  $    & ~ Index \\
\hline
(0,1,1)       &   $U(1)$   &  $U(1)$         &  $ 1+x^{2/3}+x^{8/3}-x^4+x^{14/3}+x^{16/3}-x^6+x^{22/3}-x^8 +\ldots  $  \\
\hline
(0,1,2)       &   $U(2)$   &  $U(0)$         &  $  1  $  \\
\hline
\hline
$n=2,  N_f+k=2$ & $U(N_c)$    &   $U(4 - N_c)  $    &  \\
\hline
(0,2,1)       &   $U(1)$   &  $U(3)$         &  $  1+x^{2/3}+x^{8/3}-x^4+\ldots$  \\
\hline
(1,1,1)       &   $U(1)$   &  $U(3)$         &  $  1+x^{2/3}+x^{8/3}+(1+x^{2/3}) x^{2 r}+x^{4 r}+\ldots $  \\
\hline
(0,2,2)       &   $U(2)$   &  $U(2)$         &  $  1+x^{2/3}+x^{4/3}+x^{8/3}+x^{10/3}-x^4+x^{14/3} +\ldots $  \\
\hline
(1,1,2)       &   $U(2)$   &  $U(2)$         &  $  1+x^{2/3}+x^{4/3}+x^{4 r}+x^{2 r} (1+2 x^{2/3}+x^{4/3})+x^{-2 r} (-x^{8/3}-x^{10/3})+\ldots $  \\
\hline
(1,1,3)       &   $U(3)$   &  $U(1)$         &  $  1+x^{2/3}-x^2-x^{8/3}+x^{4 r}+x^{2 r} (1+2 x^{2/3}+x^{4/3})+x^{-2 r} (-x^2-x^{8/3}-x^{10/3}) +\ldots $  \\
\hline
\hline
$n=2,  N_f+k=3$ & $U(N_c)$    &   $U(6 - N_c)  $    &  \\
\hline
(0,3,3)       &   $U(3)$   &  $U(3)$         &  $  1+x^{2/3}+x^{4/3}+x^2+x^{8/3}+2 x^{10/3}+\ldots  $  \\
\hline
(1,2,3)       &   $U(3)$   &  $U(3)$         &  $  1+x^{2/3}+x^{4/3}-x^2-3 x^{8/3}-x^{\frac{10}{3}-2 r}+x^{4 r}+x^{2 r} (1+2 x^{2/3}+2 x^{4/3}) +\ldots $  \\
\hline
(2,1,3)       &   $U(3)$   &  $U(3)$         &  $ 1+x^{2/3}+x^{4/3}-7 x^2-15 x^{8/3}-x^{4-4 r}-4 x^{\frac{10}{3}-2 r}+4 x^{3-r}+10 x^{4 r}+4 x^{\frac{7}{3}+r}+x^{2 r} (4+8 x^{2/3}+8 x^{4/3}) +\ldots  $  \\
\hline
\hline
$ n=3, N_f+k=1$ & $U(N_c)$    &   $U(3 - N_c)  $    & \\
\hline
(0,1,1)       &   $U(0)$   &  $U(3)$         &  $ 1   $  \\
\hline
(0,1,1)       &   $U(1)$   &  $U(2)$         &  $ 1+\sqrt{x}+x+x^{5/2}+x^3-x^4+2 x^5+x^{11/2}-x^6 +\ldots  $  \\
\hline
\hline
$ n=3, N_f+k=2$      & $U(N_c)$    &   $U(6 - N_c)  $                    &  \\
\hline
(0,2,3)       &   $U(3)$   &  $U(3)$         &  $ 1+\sqrt{x}+2 x+2 x^{3/2}+x^2+x^{5/2}+2 x^3 +\ldots  $  \\
\hline
(1,1,3)       &   $U(3)$   &  $U(3)$         &  $ 1+\sqrt{x}+2 x+2 x^{3/2}+x^2-x^{\frac{5}{2}-2 r}+x^{4 r}+x^{2 r} (1+2 \sqrt{x}+4 x)+\ldots   $  \\
\hline
\hline
$ n=4, N_f+k=1$      & $U(N_c)$    &   $U(4 - N_c)  $                    &  \\
\hline
(0,1,2)       &   $U(2)$   &  $U(2)$         &  $ 1+x^{2/5}+2 x^{4/5}+x^{6/5}+x^{8/5}+x^{12/5}+2 x^{14/5} +\ldots  $  \\
\hline
(0,1,3)       &   $U(3)$   &  $U(1)$         &  $ 1 + x^{2/5} + x^{4/5} +\ldots  $  \\
\hline
\caption{Superconformal index for $U(N)$ gauge theories with an adjoint and a superpotential.}
\end{longtable}

The $S^3$ partition function for rank higher than $2$ is difficult to evaluate because the integrand oscillates rapidly. Apart from $N_f=0$, the only case which we were able to test using Mathematica is $N_f=k=1,N_c=2$ and $W=\Tr X^3$. In this case both the electric and magnetic gauge groups are $U(2)$, but on the magnetic side there are singlet fields $M_j$, $j=1,2$, which couple to $q,\tilde q$ and $Y$ via a superpotential. The partition functions agree well numerically; their values as a function of $r=R(Q)$ are displayed in Table \ref{table2}.

\begin{table}[h]
\begin{tabular}{|c|ccccccc|}
\hline
$r-1/3$ & -0.3 & -0.2 & -0.1 & 0 & 0.1 & 0.2 & 0.3 \\
$\log |Z|$ & -0.423782 & -1.66927 & -1.94454 & -1.91804 & -1.73155 & -1.45191 & -1.12236\\ \hline
\end{tabular}
\caption{$S^3$ partition function for $U(2)$ theory with $N_f=k=1$, an adjoint superfield $X$, and a superpotential $W=\Tr\, X^3$ as a function of $r=R(Q)$. The dual theory is the same, plus singlets $M_1, M_2$ coupled via a superpotential. The partition functions of the electric and magnetic theories agree to the accuracy achieved.\label{table2}}
\end{table}

Next, let us look at the chiral ring structure. We already discussed the action of duality on the chiral ring generators. These generators also give identifiable contributions to the index: the generator $\Tr\, X^i$ contributes a term $x^{\frac{2i}{n+1}}$ to the index, while the matrix generator $(M_j)^a_b$ contributes the term $N_f^2\cdot x^{2r+\frac{2j}{n+1}}$. Of course, products of generators (if they are not $Q$-exact) also contribute to the index. For example, for $n=2$ the generators contribute as follows: $\Tr\, X\sim x^{2/3}$, $(M_1)^a_b \sim N_f^2 \cdot x^{2r}$ and $(M_2)^a_b \sim N_f^2 \cdot x^{2r+2/3}$ term. But $\Tr\, X \cdot(M_1)^a_b$ has the same energy as $(M_2)^a_b$. Thus for $N_c>1$ the total contribution to the index is $2N_f^2 \cdot x^{2r+2/3}$. For $N_c=1$ $(M_2)^a_b$ is identical to $\Tr\, X \cdot (M_1)^a_b$, so the total contribution is $N_f^2 \cdot x^{2r+2/3}$.

Let us now discuss the nonperturbative truncation of the chiral ring. On the classical level the relations in the chiral ring arise both from the superpotential and the characteristic equation satisfied by $X$. Thus on the classical level the independent generators among $\Tr\, X^i$ are those with $i\leq {\rm min}(n-1,N_c)$. This may lead to a conflict with duality, since in general the ranks of the electric and magnetic gauge groups are different. In the 4d case, it was proposed in \cite{KutasovSchwimmer95} that on the quantum level there are additional relations in the chiral ring of the electric (resp. magnetic) theory coming from the magnetic (resp. electric) characteristic equation. In the 3d case, we are able to check this for some dual pairs of theories.

We have seen already some examples of nonperturbative truncation for $N_f=0$ and $k=1$. For example, if $N_c=2$ and $n=3$, the electric chiral ring is spanned classically by $\Tr\, X$ and $\Tr\, X^2$. But since the magnetic gauge group is $U(1)$, the magnetic characteristic equation imposes an extra relation $\Tr\, Y^2=(\Tr\, Y)^2$, which means that there should be only one independent generator $v=\Tr\, X$ satisfying the relation $v^3=0$. To see how this comes about, recall that by virtue of the Jafferis-Yin duality the superpotential deformation $\Tr\, X^4$ can be written in terms of free superfields $v=\Tr\, X$ and $u=\Tr\, (X-v/2)^2$ as $u^2/2+3uv^2/2+v^4/8$. This leads to a chiral ring relation $u=-3v^2/2$. Thus the quantum chiral ring is generated by $v$, as predicted. The relation $v^3=0$ is also reproduced since after expressing $u$ through $v$ the superpotential becomes $W=-v^4$.

Another example of nonpertubative truncation occurs for $(n,N_f,k,N_c)=(4,0,1,3)$. The magnetic gauge group is $U(1)$, thus duality predicts that on the quantum level only $\Tr\, X$ is an independent generator. On the classical level however the operators $\Tr\, X$, $\Tr\, X^2$ and $\Tr\, X^3$ are all independent. Superconformal index suggests a mechanism for this truncation. The index on the electric side receives a contribution $2\cdot x^{4/5}$ from $(\Tr\, X)^2$ and $\Tr\, X^2$. But it also receives a contribution $- x^{4/5}$ from the monopole operator with magnetic flux $(m_1,m_2,m_3)=(-1, 0, 1)$. This suggests that on the quantum level $\Tr\, X^2$ ``pairs'' up with a monopole operator and disappears  (i.e. becomes either exact or not closed).

As mentioned above, the duality is expected to hold only for small enough $n$, namely for those $n$ for which $W=\Tr\, X^{n+1}$ is a relevant deformation of the theory with $W=0$. Let us determine this range for $N_c=2$ and several small values of $N_f$ and $k$. First of all, as discussed above, for $N_f=0$ and $k=1$ the IR dimensions of $v=\Tr \, X$ and $u=\Tr\, (X-v/2)^2$ are $1/2$. Hence the superpotential $W=\Tr \,X^{n+1}$ is relevant for $n<6$. For larger $n$ it contains only irrelevant or marginally irrelevant terms. For $N_f=0$ and $k=2$ we do not have a simple dual description of the theory with $W=0$ and so we do not know the exact scaling dimension of $u$ (the scaling dimension of $v$ is always $1/2$, because it is decoupled). However, we can determine the dimension of $u$ numerically using $F$-maximization.\footnote{This procedure is valid only if the superconformal R-symmetry is not accidental. The fact that $\log |Z|$ indeed has a minimum as a function of $r$ and this minimum obeys the unitarity bound provides some evidence that this is indeed the case.} This gives $R(u)\simeq 0.58$ \cite{JafferisYin11}. From this we conclude that the superpotential $W=\Tr X^{n+1}$ is relevant for $n<6$. Similarly, for $N_f=1$ and $k=1$ $F$-maximization gives $R(u)\simeq 0.54$, hence the superpotential $W=\Tr \, X^{n+1}$ is relevant for $n<6$. For large $N_f$ and/or $k$ the theory becomes weakly-coupled and therefore $R(u)$ approaches $1$. Thus for $N_f\gg 1$ or $k\gg 1$ the superpotential $W=\Tr\, X^{n+1}$ is relevant for $n<3$.

\subsection{$O(N)$ with an adjoint}
The Kutasov-Schwimmer-Seiberg duality in 4d has been generalized to other classical groups and types of two-index tensors in  \cite{Intriligator95, LeighStrassler95, IntriligatorLeighStrassler95} . In the rest of this paper we propose similar generalizations in 3d. In this subsection we consider replacing $U(N)$ with $O(N)$. Specifically we propose that the following two $\cN=2$ $d=3$  Chern-Simons theories are dual:

\begin{itemize}

\item Electric theory: $O(N_c)_k$ gauge theory with $N_f$ fundamental chiral multiplets $Q^a$ with $a=1,\ldots, N_f,$ and an adjoint chiral multiplet $X$ with a superpotential $W_e=\Tr\, X^{2(n+1)}$.

\item Magnetic theory: $O((2n+1)(N_f+k)+2-N_c)_{-k}$ gauge theory with $N_f$ fundamental chiral multiplets $q_a$ with  $a=1,\ldots,N_f$ , an adjoint chiral multiplet $Y$, color-singlet chiral multiplets $M_{j}^{ab}$, $j=0,\ldots,2n,$ $a,b=1,\ldots,N_f$ which are symmetric (resp. anti-symmetric) for even (resp. odd) $j$, and a superpotential $W_m=\Tr\, Y^{2(n+1)}+\sum_{j=0}^{2n} M_j^{ab} q_a Y^{2n-j} q_b.$

\end{itemize}

This is an orthogonal version of the Niarchos duality. It is supposed to hold for small enough $n$, so that the superpotential deformations $\Tr\, X^{2n+2}$ and $\Tr\, Y^{2n+2}$ are relevant.
R-charge assignments are given by $R(Q)=r, R(q)=\frac{1}{n+1}-r,
R(M_j)=\frac{j}{n+1}+2r, R(X)=R(Y)=\frac{1}{n+1}$. Note that for $n=0$ the above duality is equivalent to the duality considered in \cite{Kapustin11}.

Let us begin by considering the case $N_f=0, k=1$. The above duality says that
$O(N_c)_1$ gauge theory with an adjoint $X$ and $W=\Tr \, X^{2(n+1)}$ is dual to $O(2n+3-N_c)_{-1}$ gauge theory with an adjoint $Y$ and
$W=\Tr\, Y^{2(n+1)}$. If $N_c=2n+3$ or $N_c=2n+2$, the dual theory is
trivial (note that when the gauge group is $O(1)$, the adjoint is zero). If $2n+3<N_c$, then the rank of the magnetic gauge group is negative which we interpret as a sign that supersymmetry is spontaneously broken.

It turns out that some of these results follow naturally from an
$SO$ version of the Jafferis-Yin duality:
\begin{itemize}
\item $SO(2N_c+2)$ gauge theory with Chern-Simons level 1, an adjoint chiral superfield $X$ and $W=0$ is dual to a theory of $N_c+1$  free chiral superfields  $\sigma_{2}$,
$\sigma_4$, $\ldots$, $\sigma_{2N_c}$ and $p$.
\end{itemize}
This duality maps the chiral operator $\Tr\, X^{2j}$ to $\sigma_{2j}$ and the chiral operator ${\rm Pf} \, X$ to $p$.
Thus if we assign $X$ R-charge $r$, then $\sigma_{2j}$ should be assigned R-charge $2rj$, and $p$ should have R-charge $(N_c+1)r$. Note that this R-symmetry is not the superconformal R-symmetry. The latter would assign R-charge $1/2$ to all $\sigma_{2j}$ and $p$. The superconformal R-symmetry is an accidental symmetry from the point of view of the $SO(2N_c+2)$ gauge theory. This duality is partially motivated by the fact that the moduli space of the $SO(2N_c+2)$ gauge theory is parameterized by the expectation values of the fields $\sigma_{2j}$, $j=1,\ldots,N_c$ and $p$ which do not satisfy any constraints. Note also that at a generic point in the moduli space the low-energy theory contains, apart from moduli, $N_c+1$ copies of the $SO(2)$ Chern-Simons theory at level $1$. However, since the latter theory is essentially trivial \cite{Witten sl2z}, there is no contradiction with the conjectured duality.

An orthogonal version of the Jafferis-Yin duality is obtained from the $SO$ version by gauging a
discrete symmetry which acts by an automorphism of $SO(2N_c+2)$. Under this $\ZZ_2$ symmetry the
fields $\sigma_{2j}=\Tr\, X^{2j}$ are even, while ${\rm Pf}\, X$ is odd. Thus the dual of $O(2N_c+2)_1$ gauge theory
with an adjoint is a $\ZZ_2$ orbifold of the free theory of superfields $\sigma_{2j}$ and $p$ where the generator
of $\ZZ_2$ flips the sign of $p$ and leaves all other superfields invariant.

Let us perturb both sides of the orthogonal Jafferis-Yin duality by a superpotential $W=\Tr\, X^{2(N_c+1)}$. Thanks to
the characteristic equation satisfied by $X$ and the identity $\det \, X={(\rm Pf}\, X)^2$, this superpotential is
a quadratic function of the fields $p,\sigma_{2j}$ with a unique nondegenerate critical point $p=\sigma_{2j}=0$.
Hence the IR limit of the perturbed theory is trivial as expected. If instead we perturb the theory
by $W=\Tr\, X^{2(n+1)}$ with $n=0,1,\ldots,N_c-1$, the symmetry is broken spontaneously, since in terms of
dual free fields such a superpotential is a linear function. This provides some evidence for the orthogonal version
of the Jafferis-Yin duality.

As a further check let us verify that for small $N_c$ the $SO$
version of the Jafferis-Yin duality follows from other known
results. For $N_c=0$ the electric theory becomes a product of a free
theory of a single chiral superfield $X$ and $U(1)$ Chern-Simons
theory at level $1$. Since $U(1)_1$ Chern-Simons  theory is
essentially trivial \cite{Witten sl2z}, this agrees with the duality
statement. For $N_c=1$ the electric theory has gauge group
$SO(4)=(SU(2)\times SU(2))/\ZZ_2$, and the adjoint of $SO(4)$ is a
sum of adjoints  for  each $SU(2)$ factor. Hence by virtue of the
``Duality Appetizer" duality $SO(4)_1$ theory with an adjoint $X$ is
dual to a theory of two free chiral fields $\sigma_2$ and
$\sigma'_{2}$. For $N_c=2$ the electric gauge group is
$SO(6)=SU(4)/\ZZ_2$, so the statement of the $SO$ duality is
essentially the same as the statement of the unitary Jafferis-Yin
duality discussed in the previous section.

The first case of $SO$ Jafferis-Yin duality which does not reduce to any previously-studied duality is when the
electric gauge group is $SO(8)$. To test it, we computed the superconformal index of $SO(8)_1$ and $O(8)_1$
theories with an adjoint and $W=0$, the results are displayed below:

\begin{enumerate}
%







\item $SO(8)_1$ with one adjoint X of R-charge  $r$ $\thickapprox$ Four free chiral multiplets $\sigma_{2}$, $\sigma_4$,
 $\sigma_6$ and $p$ with R-charges $2r$, $4r$, $6r$, $4r$.

$I = 1-11 x^2-x^{2-6 r}-3 x^{2-4 r}-6 x^{2-2 r}+x^{2 r}+\ldots$

\item  $O(8)_1$  one adjoint X of R charge $r$  $\thickapprox$ A $\ZZ_2$-orbifold of the theory of four free chiral multiplets $\sigma_{2}$, $\sigma_4$,
$\sigma_6$ and $p$ with R-charges $2r$, $4r$, $6r$, $4r$.

$I = 1-7 x^2-x^{2-6 r}-2 x^{2-4 r}-4 x^{2-2 r}+x^{2 r}+\ldots$

\end{enumerate}

For $N_f>1$ or $k>1$ we do not have a simple dual description of the theory without the superpotential. For theories with the superpotential $W=\Tr\, X^{2(n+1)}$ we worked out the superconformal index for several low values of $N_f,N_c,k$ and $n$. The results are displayed in Table \ref{table3} and are in complete agreement with the duality predictions.
\begin{longtable}{|c|c|c|p{8cm}|}
\hline
$(N_f,k,N_c)$& Electric    &   Magnetic     &  \\
$n=1, N_f+k=1$ & $O(N_c)$    & $O(5-N_c)$    & ~ Index \\
\hline
(0,1,1)       &   $O(1)$  &  $O(4)$         &  $  1  $  \\
\hline
(0,1,2)       &   $O(2)$  &  $O(3)$         &  $  1+x+x^3-x^4 +\ldots $  \\
\hline
\hline
$n=1, N_f+k=2$&  $O(N_c)$ &   $O(8-N_c)$    &  \\
\hline
(0,2,2)       &   $O(2)$   &  $O(6)$         &  $ 1 + x + x^3 +\ldots  $  \\
\hline
(1,1,2)       &   $O(2)$   &  $O(6)$         &  $ 1+x+x^3+x^{6 r}+x^{2 r} (1+x)+x^{4 r} (1+x)+\ldots  $  \\
\hline
(0,2,3)       &   $O(3)$   &  $O(5)$         &  $  1+x-x^2+2 x^3+2 x^{7/2}-2 x^4 +\ldots $  \\
\hline
(1,1,3)       &   $O(3)$   &  $O(5)$         &  $  1+x-x^3+x^{6 r}+x^{4 r} (1+2 x)+x^{2 r} (1+2 x+x^2) +\ldots $  \\
\hline
(0,2,4)       &   $O(4)$   &  $O(4)$         &  $ 1+x+x^2+2 x^3+x^5 +\ldots $  \\
\hline
(1,1,4)       &   $O(4)$   &  $O(4)$         &  $  1+x+x^2-x^3-x^{3-2 r}+x^{6 r}+x^{4 r} (1+2 x)+x^{2 r} (1+2 x+2 x^2) +\ldots $  \\
\hline
(1,1,5)       &   $O(5)$   &  $O(3)$         &  $  1+x-2 x^3-5 x^4+x^{8 r}+x^{6 r} (1+2 x)+x^{4 r} (1+2 x+3 x^2)+x^{2 r} (1+2 x+2 x^2-x^3)+x^{-2 r} (-x^3-x^4)+\ldots $  \\
\hline
(1,1,6)       &   $O(6)$   &  $O(2)$         &  $ 1+x-x^2-2 x^3+x^{6 r}+x^{4 r} (1+2 x)+x^{2 r} (1+2 x+x^2)+x^{-2 r} (-x^2-2 x^3)+\ldots  $  \\
\hline
\hline
$n=1, N_f+k=3$&  $O(N_c)$ &   $O(11-N_c)$    &  \\
\hline
(2,1,5)       &   $O(5)$   &  $O(6)$         &  $ 1+x-3 x^2+6 x^{4 r}+x^{2 r} (3+\sqrt{x}+6 x) +\ldots  $  \\
\hline
\hline
$n=2, N_f+k=1$&  $O(N_c)$ &   $O(7-N_c)$    &  \\
\hline
(0,1,1)       &   $O(1)$   &  $O(6)$         &  $  1  $  \\
\hline
(0,1,2)       &   $O(2)$   &  $O(5)$         &  $  1+x^{2/3}+x^{4/3}+x^{8/3} +\ldots $  \\
\hline
(0,1,3)       &   $O(3)$   &  $O(4)$         &  $  1+x^{2/3}+x^{4/3}+x^{8/3}+\ldots  $  \\
\hline
\hline
$n=2, N_f+k=2$&  $O(N_c)$ &   $O(12-N_c)$    &  \\
\hline
(0,2,6)       &   $O(6)$   &  $O(6)$         &  $ 1+x^{2/3}+2 x^{4/3}+2 x^2+\ldots   $  \\
\hline
(1,1,6)       &   $O(6)$   &  $O(6)$         &  $1+x^{2/3}+2 x^{4/3}+2 x^2+x^{8/3}+(1+2 x^{2/3}) x^{6 r}+x^{8 r}+x^{4 r} (1+2 x^{2/3}+5 x^{4/3})+x^{2 r} (1+2 x^{2/3}+4 x^{4/3}+5 x^2)+x^{-2 r} (-x^{8/3}-2 x^{10/3})+\ldots   $  \\
\hline
\caption{Superconformal index for $O(N)$ gauge theories with an adjoint and a superpotential.\label{table3}}
\end{longtable}

In the case $N_c=3,k=1,N_f=1,n=1$ we also computed the $S^3$ partition function of both electric and magnetic theories and verified that they agree up to a phase. The results of this computation are presented in Table \ref{table4}. Note that for $N_f=1$ the singlet fields $M_j$ on the magnetic side exist only for even $j$ (one for each even $j$, $j=0,\ldots, 2n$).

\begin{table}[h]
\begin{tabular}{|c|ccccccc|}\hline
$r$ & 0.10 & 0.15 & 0.20 & 0.25 & 0.30 & 0.35 & 0.40 \\
$\log |Z| $ & -1.49400 & -1.82653 & -2.01289 & -2.11015 & -2.14535 & -2.13404 & -2.08640 \\ \hline
\end{tabular}
\caption{$S^3$ partition function of the $O(3)_1$ theory with one fundamental superfield $Q$, an adjoint $X$, and $W=\Tr\, X^4,$ as a function of $r=R(Q)$. The magnetic dual theory has the same partition function to the accuracy achieved.\label{table4}}
\end{table}

%

\subsection{$Sp(2N)$ with an antisymmetric tensor}

In this section we consider a duality for $Sp(2N_c)$ gauge theories
with $2N_f$ fundamental superfields and an antisymmetric tensor
superfield. \footnote{Our conventions are that $Sp(2N)$ has rank $N$
so that $Sp(2)=SU(2)$. The number of fundamental superfields is
taken to be even so that the Chern-Simons coupling is integral. If
the number of fundamentals were odd, cancelation of global anomalies
would require the Chern-Simons coupling to be half-integral.} This
theory was also discussed in \cite{Vartanov11}. Let $J$ be the
invariant antisymmetric tensor in the product of the
$2N$-dimensional representation of $Sp(2N)$ with itself. An
antisymmetric tensor $X$ satisfies $X^t=-X$. Note that $\Tr\, (XJ)$
is gauge-invariant, and therefore the antisymmetric tensor
representation is reducible: it decomposes into a singlet and the
antisymmetric traceless tensor representation. The latter
representation is irreducible. The singlet part couples to the rest
of the theory only through the superpotential. A slightly different
theory is obtained by requiring $X$ to satisfy $\Tr (XJ)=0$. All the
results described below can be easily modified to accommodate this
difference.

We propose the following dual pair:

\begin{itemize}

 \item Electric: $Sp(2N_c)_k$ gauge theory with $2N_f$ fundamental chiral multiplets $Q^a$,  $a=1,\ldots, 2N_f$, an antisymmetric chiral mutliplet $X$, and a  superpotential $W_e=\Tr\, (XJ)^{n+1};$

\item Magnetic: $Sp(2\big(n(N_f+k-1)-N_c\big))_{-k}$ gauge theory with $2N_f$ fundamental chiral multiplets $q_a$, $a=1,\ldots, 2N_f$, an antisymmetric chiral multiplet $Y$, and $n\cdot N_f(2N_f-1)$ singlet chiral multiplets $M_{j}^{ab}$ , $j=1,\ldots,n$ with the superpotential
$$W_m=\Tr\, (YJ)^{n+1}+\sum_{j=1}^{n} M^{ab}_j q_a J (YJ)^{n-j} q_b.
$$

\end{itemize}

This duality is supposed to hold for small enough $n$, so that the superpotential deformations
$\Tr\, (XJ)^{n+1}$ and $\Tr\, (YJ)^{n+1}$ are relevant.
R-charge assignments are  $R(Q)=r, R(q)=\frac{2}{n+1}-r, R(M_j)=\frac{2(j-1)}{n+1}+2r, R(X)=R(Y)=\frac{2}{n+1}$.
The duality maps the chiral ring generators $\Tr\, (XJ)^j$ to $\Tr (YJ)^j$, $j=1,\ldots,n-1$  and
the chiral ring generators $Q^a J (XJ)^{j-1} Q^b$ to $M_j^{ab}$, $j=1,\ldots,n$.

As before we begin with the case $N_f=0$.
If we set the Chern-Simons level $k$ to $2$, then $Sp(2N_c)_2$ theory with
$W=\Tr\, (XJ)^{n+1}$ is supposed to be dual to $ Sp(2(n-N_c))_2$ theory with $W=\Tr\,
(YJ)^{n+1}$. Letting $n=N_c$, we are led to the conclusion that
$Sp(2N_c)_2$ theory with $W=\Tr\, (XJ)^{N_c+1}$ is a trivial theory. Letting $n<N_c$ we conclude that
$Sp(2N_c)_2$ theory with $W=\Tr\, (XJ)^{n+1}$ for $n<N_c$ breaks supersymmetry spontaneously.
This behavior suggests the following symplectic version of the Jafferis-Yin duality:

\begin{itemize}

\item $Sp(2N_c)$ gauge theory with Chern-Simons level $k=2$, an antisymmetric tensor $X$ with
R-charge $r$ and zero superpotential is dual to a theory of $N_c$ free chiral multiplets with R-charges
 $r, 2r,\ldots, N_c r$.

\end{itemize}

The simplest check of this duality is to note that at a generic
point in the moduli space the gauge group is broken down to
$Sp(2)^{N_c}$, where each $Sp(2)=SU(2)$ factor has Chern-Simons
level $2$. Apart from this TQFT, the only other fields in the
low-energy theory are the moduli with R-charges $r$, $2r,\ldots, N_c
r$ which correspond to the chiral ring generators $\Tr\, (XJ)^j$ for
$j=1,\ldots,N_c$. Now we note that $SU(2)_2$ Chern-Simons theory is
trivial,\footnote{Here it is important to remember about the
difference between the supersymmetric $SU(2)$ Chern-Simons theory
and bosonic $SU(2)$ Chern-Simons theory: the former theory at level
$k$ is equivalent to the latter theory at level $k-2\, \sign(k)$
\cite{Wittenindex}. Hence supersymmetric Chern-Simons at level $2$
is equivalent to bosonic Chern-Simons at level $0$, which is
trivial.} therefore along the moduli space the duality holds true.

Another simple check is to note that $W=\Tr\, (XJ)^{N_c+1}$ is a massive superpotential when written in terms of free
fields $\sigma_j=\Tr\, (XJ)^j$, $j=1,\ldots,N_c$. Hence this deformation leads to a trivial theory. On the other hand,
if we deform the theory by adding a superpotential $W=\Tr\, (XJ)^j$ for $j=1,\ldots,N_c$, we get spontaneous supersymmetry breaking, in agreement with the duality conjecture.

To test this duality further,  one can compute the superconformal index of the $Sp(2N_c)_2$ theories with
an antisymmetric tensor superfield $X$ . Below we list the results for a couple low values of $N_c$.

\begin{enumerate}

\item $Sp(2)_2$ gauge theory with an antisymmetric $X $of R-charge $r$  $\thickapprox$ One free chiral
multiplet with R-charge
$r$

$I=1-x^2-2 x^4+x^r+x^{2 r}+x^{3 r}+x^{4 r}+x^{5 r}+x^{6 r}+x^{7 r}+x^{8 r}+x^{-r} (-x^2-x^4)$+\ldots

\item  $Sp(4)_2$ gauge theory with an antisymmetric $X$ of R-charge $r$  $\thickapprox$ Two free chiral
multiplets with R-charge
$r$, $2r$

$I = 1-x^2-2 x^4+x^r+x^{2 r}+x^{3 r}+x^{4 r}+x^{5 r}+x^{6 r}+x^{7 r}+x^{8 r}+x^{-r} (-x^2-x^4)+\ldots $

\item $Sp(6)_2$ with an antisymmetric $X$ of R-charge $r$ $\thickapprox$ Three free chiral multiplets with R-charge $r$,
$2r$, $3r$

$I =1-6 x^2-x^{2-3 r}-2 x^{2-2 r}-4 x^{2-r}+2 x^{2 r}+3 x^{3 r}+x^r (1-8 x^2)+\ldots$

\end{enumerate}

The first of these dualities is obvious, since for $Sp(2)$  gauge group $X$ is a free field, and $Sp(2)_2$ $\cN=2$ Chern-Simons theory is trivial. The second one is not really new either: $Sp(4)_2$ theory with an antisymmetric $X$ is equivalent to the $O(5)_2$ theory with $N_f=1$ matter superfield in the vector representation plus a decoupled free field (the ``trace'' part of $X$).\footnote{One might naively think that $Sp(4)_2$ theory is related to $SO(5)_2$ theory. But since $SO(5)=Sp(4)/\ZZ_2$, the two theories have different magnetic flux quantization conditions, and as a consequence $SO(5)_2$ theory has additional monopole operators with nontrivial 't Hooft magnetic flux not allowed for $Sp(4)_2$. One can show that at Chern-Simons level $2$ these additional operators are projected out by the $\ZZ_2$ subgroup of $O(5)\simeq SO(5)\times\ZZ_2$. So $Sp(4)_2$ theory is equivalent to $O(5)_2$ theory.} But $O(5)_2$ theory with a single vector is equivalent to a free chiral superfield by virtue of the orthogonal Seiberg duality of \cite{Kapustin11}. Hence $Sp(4)_2$ with an antisymmetric $X$ is equivalent to a pair of free superfields. The third duality is genuinely new.

For $N_f>0$ we check the symplectic duality with $W=\Tr\, (XJ)^{n+1}$ by computing the superconformal index for several low values of $N_f, N_c, k, n$. The indices of dual theories agree in all the cases we studied.

\begin{longtable}{|c|c|c|p{8cm}|}
\hline
$(N_f,k,N_c)$& Electric    &   Magnetic     &  \\
$n=2, N_f+k=2$ & $Sp(2N_c)$    & $Sp(2(2-N_c))$    & ~ Index \\
\hline
(0,2,1)       &   $Sp(2)$  &  $Sp(2)$         &  $  1+x^{2/3}+x^{8/3}+\ldots  $  \\
\hline
(1,1,1)       &   $Sp(2)$  &  $Sp(2)$         &  $ 1+x^{2/3}-x^2+(1+x^{2/3}) x^{2 r}+x^{4 r}+x^{-2 r} (-x^2-x^{8/3}) +\ldots $  \\
\hline
\hline
$n=2, N_f+k=3$ & $Sp(2N_c)$    & $Sp(2(4-N_c))$ & \\
\hline
(0,3,1)       &   $Sp(2)$  &  $Sp(6)$         &  $  1+x^{2/3}+x^{8/3}+\ldots $  \\
\hline
(1,2,1)       &   $Sp(2)$  &  $Sp(6)$         &  $ 1+x^{2/3}-4 x^2-3 x^{8/3}+(1+x^{2/3}) x^{2 r}+x^{4 r} +\ldots $  \\
\hline
(2,1,1)       &   $Sp(2)$  &  $Sp(6)$         &  $ 1+x^{2/3}-16 x^2-15 x^{8/3}-x^{4-4 r}+(6+6 x^{2/3}) x^{2 r}+20 x^{4 r} +\ldots  $  \\
\hline
(0,3,2)       &   $Sp(4)$  &  $Sp(4)$         &   $ 1+x^{2/3}+x^{4/3}+x^{8/3} +\ldots  $  \\
\hline
(1,2,2)       &   $Sp(4)$  &  $Sp(4)$         &  $ 1+x^{2/3}+x^{4/3}-4 x^2-4 x^{8/3}+x^{4 r}+x^{2 r} (1+2 x^{2/3}+x^{4/3})+x^{-2 r} (-x^{8/3}-x^{10/3}) +\ldots $  \\
\hline
(2,1,3)       &   $Sp(4)$  &  $Sp(4)$         &  $ 1+x^{2/3}+x^{4/3}-16 x^2-51 x^{8/3}+21 x^{4 r}+x^{2 r} (6+12 x^{2/3}+6 x^{4/3})+x^{-2 r} (-6 x^{8/3}-12 x^{10/3})+x^{-4 r} (-x^{10/3}-x^4)+\ldots  $  \\
\hline
(1,2,3)       &   $Sp(6)$  &  $Sp(2)$         &  $ 1+x^{2/3}-5 x^2-5 x^{8/3}+x^{4 r}+x^{2 r} (1+2 x^{2/3}+x^{4/3})+x^{-2 r} (-x^2-x^{8/3}+3 x^{10/3})+\ldots  $  \\
\hline
(2,1,3)       &   $Sp(6)$  &  $Sp(2)$         &  $1+x^{2/3}-52 x^2-108 x^{8/3}+6 x^{\frac{14}{3}-6 r}+21 x^{4 r}+x^{2 r} (6+12 x^{2/3}+6 x^{4/3})+x^{-2 r} (-6 x^2-12 x^{8/3}+14 x^{10/3})+x^{-4 r} (-x^{8/3}-x^{10/3}+15 x^4)+\ldots $  \\
\hline
\hline
$n=3, N_f+k=2$ & $Sp(2N_c)$    & $Sp(2(3-N_c))$ & \\
\hline
(0,2,1)       &   $Sp(2)$  &  $Sp(4)$         &  $  1+\sqrt{x}+x+x^{5/2}+x^3+\ldots  $  \\
\hline
(1,1,1)       &   $Sp(2)$  &  $Sp(4)$         &  $  1+\sqrt{x}+x-x^2+x^{6 r}+x^{2 r} (1+\sqrt{x}+x)+x^{4 r} (1+\sqrt{x}+x)+x^{-2 r} (-x^2-x^{5/2}-x^3) +\ldots $  \\
\hline
(1,1,2)       &   $Sp(4)$  &  $Sp(2)$         &  $  1+\sqrt{x}+x-x^{3/2}-3 x^2-3 x^{5/2}-2 x^3+x^{6 r}+x^{4 r} (1+2 \sqrt{x}+3 x)+x^{2 r} (1+2 \sqrt{x}+2 x-2 x^2)+x^{-2 r} (-x^{3/2}-2 x^2-2 x^{5/2}-x^3)+x^{-4 r} (x^{7/2}+x^4) +\ldots $  \\
\hline
\caption{Superconformal index for $Sp(2N)$ gauge theories with an antisymmetric tensor and a superpotential.}
\end{longtable}

Chiral ring generators give readily identifiable contributions to
the index. For example, for $n=2$ (i.e. cubic superpotential) the
generator $\Tr (XJ)$ contributes $x^{2/3}$, while both $\Tr\,
(XJ)^{2}$ as well as $(\Tr\, (XJ))^2$ contribute a term $1\cdot
x^{4/3}$. The mesons contribute as follows: $M_{1}^{ab}\sim
N_f(2N_f-1)\cdot x^{2r}$, $M_{2}^{ab}\sim N_f(2N_f-1)\cdot
x^{2r+\frac{2}{3}}$. One can also observe the nonperturbative
truncation of the chiral ring for $Sp(6)$-$Sp(2)$ and
$Sp(4)$-$Sp(2)$ dual pairs. Consider first $n=2$ and the
$Sp(6)_e$-$Sp(2)_m$ dual pair. On the magnetic side we have a
``classical'' contribution $x^{2/3}$ from $\Tr\, (YJ)$ and no
contribution  of the form $x^{4/3}$ because for $Sp(2)$ theory
$(\Tr\, (YJ))^2$ is proportional to $\Tr\, (YJ)^2$, and the latter
is zero because of the superpotential $W_m=\Tr\, (YJ)^3$. On the
electric side, both $(\Tr\, (XJ))^2$ and $\Tr\, (XJ)^2$ contribute
to the coefficient of the term $x^{4/3}$, but this ``classical''
contribution is canceled by a monopole operator with the magnetic
charge $(1,0,0)$. Similarly, on the magnetic side there is a
contribution $x^{2r+2/3}$ from $\Tr\, (YJ)\cdot M_1$, while on the
electric side there is a classical contribution $2 x^{2r+2/3}$ from
$\Tr\, (XJ) QJQ $ and $QJXJQ$ which is partially canceled by a
monopole operator.

Another case of nonperturbative truncation occurs for $n=3$ and the
 $Sp(4)_e$-$Sp(2)_m$ dual pair. The operators $\Tr\, (XJ)$ and
$\Tr\, (YJ)$ contribute $x^{1/2}$ on both sides of the duality. On
the magnetic side the contribution $x$ comes entirely from $(\Tr\,
(YJ))^2$. On the electric side there are classical contributions of
this form from both $(\Tr\, (XJ))^2$ and $\Tr\, (XJ)^2$, but they
are partially canceled by a monopole operator with magnetic charge
$(1,0)$. Similarly, the contribution $x^{3/2}$ on the electric side
from $\Tr\, (XJ)^2\Tr\, (XJ)$ is canceled by a monopole operator
with charge $(1,0)$. Similarly, on the magnetic side there is a term
$x^{2r+1/2}$ coming from $M_1 \Tr\, (YJ) $, while on the electric
side there are contributions of this form both from $\Tr\, (XJ)\cdot
QJQ$ and $Q J X J Q$ which are partially canceled by a monopole
operator.

\subsection{$U(N)$ with an antisymmetric tensor}

An antisymmetric tensor flavor of $U(N)$ is a pair of chiral multiplets $X$ and $\tX$, one in the antisymmetric tensor representation and one in the dual representation. We propose the following duality:

\begin{itemize}

 \item Electric theory: $U(N_c)_k$ theory with $N_f$ fundamental flavors $Q^a$, $\tilde{Q}_a$, an antisymmetric tensor flavor $X$, $\tilde{X}$, and the superpotential $W_e=\Tr\, (\tilde{X}X)^{n+1}$

\item Magnetic theory: $U( (2n+1)(N_f + k) -2n - N_c)_{-k}$ theory with  $N_f$ fundamental flavors $q_a$, $\tilde{q}^a$, an antisymmetric tensor flavor $Y$, $\tilde{Y}$ , $(n+1)\cdot N_f^2$ singlets $(M_{j})^{a}_{b}$, $j=0,\cdots,n$, $n N_f(N_f-1)$ singlets $(P_l)^{ab}$ and $(\tilde{P}_l)_{ab}$, $l=0,\cdots,n-1$, and the superpotential
$$W_m=\textrm{tr}(\tilde{Y}Y)^{n+1}+\sum_{j=0}^{n} M_j q (\tilde{Y}Y)^{n-j} \tilde{q} + \sum_{l=0}^{n-1}\big( P_l q (\tilde{Y}Y)^{n-1-l}\tilde{Y}q + \tilde{P}_l \tilde{q} Y(\tilde{Y}Y)^{n-1-l} \tilde{q}\big)$$

\end{itemize}

The singlets $P_l^{ab}$ and $({\tilde P}_l)_{ab}$ are antisymmetric in flavor indices. Chiral ring generators are mapped as follows:
\begin{eqnarray}
Q^a (\tilde{X}X)^{j} \tilde{Q}_b &\ra & (M_{j})^{a}_{b},\\
Q^a (\tilde{X}X)^{l}\tilde{X}Q^b &\ra & (P_l)^{ab},\\
\tilde{Q}_a X(\tilde{X}X)^{l} \tilde{Q}_b &\ra & (\tilde{P}_l)_{ab},\\
\Tr\, (X\tilde X)^j &\ra & \Tr (Y \tilde Y)^j.
\end{eqnarray}
R-charge assignments are $R(Q)=r, R(q)=\frac{1}{n+1}-r,
R(M_j)=\frac{2j}{n+1}+2r, R(P_l)=R( \tilde{P}_l)=\frac{2l+1}{n+1}+2r,
R(X)=R(\tilde{X})=R(Y)=R(\tilde{Y})=\frac{1}{n+1}$.

We have computed the superconformal index for several dual pairs, the results are displayed in Table \ref{tb-index-UAsym}. In all cases the duality is confirmed.
\begin{longtable}{|c|c|c|p{8cm}|}
\hline
$(N_f,k,N_c)$       & Electric    &   Magnetic                    &  \\
$n=1,  N_f+k=2$ & $U(N_c)$    &   $U(4 - N_c)  $    & ~ Index \\
\hline
(0,2,1)       &   $U(1)$   &  $U(3)$         &  $  1  $  \\
\hline
(1,1,1)       &   $U(1)$   &  $U(3)$         &  $  1-x^4-x^{4-2 r}+x^{2 r}+x^{4 r}+x^{6 r}+x^{8 r} +\ldots $  \\
\hline
(0,2,2)       &   $U(2)$   &  $U(2)$         &  $  1+x-3 x^2+4 x^3-4 x^4 +\ldots $  \\
\hline
(1,1,2)       &   $U(2)$   &  $U(2)$         &  $  1+x-x^3-3 x^4+x^{8 r}+x^{6 r} (1+x)+x^{2 r} (1+x+x^2)+x^{4 r} (1+x+x^2)+x^{-2 r} (-x^2-x^3-2 x^4) +\ldots $  \\
\hline
(1,1,3)       &   $U(3)$   &  $U(1)$         &  $  1-x^2-2 x^3+x^{6 r}+x^{2 r} (1+x)+x^{4 r} (1+x)+x^{-2 r} (-x^2-x^3)+\ldots  $  \\
\hline
\hline
$n=1,  N_f+k=3$ & $U(N_c)$    &   $U(7 - N_c)  $    &  \\
\hline
(2,1,4)       &   $U(4)$   &  $U(3)$         &  $ 1+x-9 x^2-16 x^{5/2}-36 x^3-x^{3-4 r}+4 x^{3-r}+20 x^{6 r}+4 x^{\frac{5}{2}+r}+x^{4 r} (10+8 \sqrt{x}+29 x)+x^{2 r} (4+2 \sqrt{x}+8 x+2 x^{3/2}-32 x^2)+x^{-2 r} (-2 x^{5/2}-8 x^3-2 x^{7/2}) +\ldots  $  \\
\hline
\hline
$n=2,  N_f+k=2$ & $U(N_c)$    &   $U(6 - N_c)  $    &  \\
\hline
(0,2,3)       &   $U(3)$   &  $U(3)$         &  $ 1+x^{2/3}+x^{8/3} +\ldots  $  \\
\hline
(1,1,3)       &   $U(3)$   &  $U(3)$         &  $ 1+x^{2/3}-x^2-x^{8/3}+(1+2 x^{2/3}) x^{6 r}+x^{8 r}+x^{2 r} (1+2 x^{2/3}+x^{4/3})+x^{4 r} (1+2 x^{2/3}+2 x^{4/3})+x^{-2 r} (-x^2-x^{8/3}-x^{10/3}) +\ldots  $  \\
\hline
\caption{Superconformal index for $U(N)$ gauge theories with an antisymmetric tensor flavor and a superpotential.\label{tb-index-UAsym}}
\end{longtable}

It is difficult to compute the $S^3$ partition function numerically
if the rank of the gauge group is higher than two. We computed the
partition function as a function of $r$ in the case $n=1,
(N_f,k,N_c)=(1,1,2)$ where both the electric and magnetic gauge
groups are $U(2)$ and verified that they agree up to a phase. The
results are presented in Table \ref{tb-ptf-UAsym}.

\begin{table}[h]
\begin{tabular}{|c|ccccccc|}
\hline
$r$ & 0.1  &  0.15 & 0.2  & 0.25 & 0.3 & 0.35 & 0.4  \\
${\rm log} |Z|$  & -1.62436 &  -1.86564 & -1.97182 & -2.00027 &
-1.97853 &  -1.92297 & -1.84478 \\ \hline
\end{tabular}
\caption{$S^3$ partition function of  $U(2)$ theory with an
antisymmetric tensor flavor, $N_f=k=1$ and a superpotential $W=\Tr\,
(X\tilde X)^2$ as a function of $r=R(Q)$. The dual theory is the
same plus singlets $M_1$ and $M_2$ coupled via a superpotential.
The partition functions agree to the accuracy
achieved.\label{tb-ptf-UAsym} }
\end{table}

Using these results we can investigate nonperturbative truncation of
the chiral ring. On the classical level, the superpotential
truncates the chiral ring so that monomials $X(\tilde{X}X)^n$ or
$(\tilde{X}X)^n\tilde{X}$ and all operators containing them are
$Q$-exact. On electric side, the classical chiral ring contains the
following generators:  $\Tr\, (X\tilde{X})^i$($\sim 1\cdot
x^{\frac{2i}{n+1}}$ in index), $i=1,\cdots,n$, $Q^a (\tilde{X}X)^{j}
\tilde{Q}_b$ ($\sim N_f^2\cdot x^{2r+\frac{2j}{n+1}}$),
$j=0,\cdots,n$, $Q^a (\tilde{X}X)^{l}\tilde{X}Q^b$ and $\tilde{Q}_a
X(\tilde{X}X)^{l} \tilde{Q}_b$($\sim \frac{N_f(N_f-1)}{2}\cdot
x^{2r+\frac{2l+1}{n+1}}$), $l=0,\cdots,n-1$. Similarly, on the
magnetic side, the classical chiral ring contains the following
generators: $\Tr\, (Y\tilde{Y})^i$($\sim 1\cdot x^{\frac{2i}{n+1}}$
in index), $i=1,\cdots,n$, $(M_j)^a_b$($\sim N_f^2\cdot
x^{2r+\frac{2j}{n+1}}$), $j=0,\cdots,n$, $(P_l)^{ab}$ and
$(\tilde{P}_l)_{ab}$($\sim \frac{N_f(N_f-1)}{2}\cdot
x^{2r+\frac{2l+1}{n+1}}$), $l=0,\cdots,n-1$. The generators appear
to match. However, for low enough rank of the electric or magnetic
gauge group additional relations appear on the classical level which
may require new nonperturbative relations in the chiral ring of the
dual theory.

Consider the following example. Let $(n,N_f,k,N_c)=(1,1,1,3)$, so that the electric gauge group is $U(3)_1$, the magnetic gauge group is $U(1)_{-1}$, and the electric superpotential is quartic, $W=\Tr\, (X\tilde X)^2$. For $N_c>1$ we have a chiral ring generator $\Tr\, X\tilde X$ which contributes a term $1\cdot x$ to the index. However, since the magnetic gauge group is $U(1)$, in the magnetic theory the antisymmetric tensor vanishes, and there is no operator which gives such a contribution. On the electric side the contribution of $\Tr\, X\tilde X$ is canceled by a monopole operator. Thus $\Tr\, (X\tilde X)$ is truncated by a quantum effect on the electric side. Further, for generic $N_c$ on the electric side we expect contributions of the form $x^{2r+1}$ from $Q {\tilde X} X \tilde Q$ and $Q\tilde Q\cdot\Tr \, X\tilde X$, so its coefficient is $2N_f^2=2$. However, if $\Tr\, X\tilde X$ is truncated, the coefficient of $x^{2r+1}$ must be $N_f^2=1$, which is indeed what we see in this case.

\subsection{$U(N)$ with a symmetric tensor}
A symmetric tensor flavor of $U(N)$ is a pair of chiral multiplets $X$ and $\tX$, one in the symmetric tensor representation and one in the dual representation. We propose the following duality:

\begin{itemize}

 \item Electric theory: $U(N_c)_k$ gauge theory with $N_f$ fundamental flavors $Q^a$, $\tilde{Q}_a$, a symmetric tensor flavor $X$, $\tilde{X}$ and a superpotential $W_e=\textrm{tr}(\tilde{X}X)^{n+1}.$

\item Magnetic theory: $U( (2n+1)(N_f + k) +2n - N_c)_{-k}$ gauge theory with  $N_f$ fundamental flavors $q_a$, $\tilde{q}^a$,  a symmetric flavor $Y$, $\tilde{Y}$,  $(n+1)N_f \times N_f$ singlets $(M_{j})^{a}_{b}$, $j=0,\cdots,n$, $nN_f(N_f+1)$ singlets $(P_l)^{ab}$, $(\tilde{P}_l)_{ab}$, $l=0,\cdots,n-1$, and a superpotential
$$
W_m=\Tr\, (\tilde{Y}Y)^{n+1}+\sum_{j=0}^{n} M_j q (\tilde{Y}Y)^{n-j} \tilde{q} + \sum_{l=0}^{n-1}\big( P_l q (\tilde{Y}Y)^{n-1-l}\tilde{Y}q +  \tilde{P}_l \tilde{q} Y(\tilde{Y}Y)^{n-1-l} \tilde{q}\big)
$$

\end{itemize}

Chiral ring generators map as follows:
\begin{eqnarray}
Q^a (\tilde{X}X)^{j} \tilde{Q}_b & \ra & (M_{j})^{a}_{b},\\
 Q^a (\tilde{X}X)^{l}\tilde{X}Q^b & \ra & (P_l)^{ab},\\
 \tilde{Q}_a X(\tilde{X}X)^{l} \tilde{Q}_b & \ra &  (\tilde{P}_l)_{ab},\\
 \Tr\, (X\tilde X)^j & \ra & \Tr\, (Y\tilde Y)^j.
\end{eqnarray}
R-charge assignments are $R(Q)=r, R(q)=\frac{1}{n+1}-r, R(M_j)=\frac{2j}{n+1}+2r,
R(P_l)=R( \tilde{P}_l)=\frac{2l+1}{n+1}+2r, R(X)=R(\tilde{X})=R(Y)=R(\tilde{Y})=\frac{1}{n+1}$.

We have computed the superconformal index for several dual pairs, the results are displayed in Table \ref{tb-index-USym}.
In all cases the duality is confirmed.\\
\begin{longtable}{|c|c|c|p{8cm}|}
\hline
$(N_f,k,N_c)$   & Electric    &   Magnetic                    &  \\
$n=1,  N_f+k=1$ & $U(N_c)$    &   $U(5 - N_c)  $    & ~ Index \\
\hline
(0,1,1)       &   $U(1)$   &  $U(4)$         &  $  1+x-x^2 +\ldots $  \\
\hline
(0,1,2)       &   $U(2)$   &  $U(3)$         &  $  1+x-x^2+2 x^3-x^4 +\ldots $  \\
\hline
$n=1,  N_f+k=2$ & $U(N_c)$    &   $U(8 - N_c)  $    &  \\
\hline
(1,1,4)       &   $U(4)$   &  $U(4)$         &  $ 1+x-2 x^2-2 x^{5/2}-3 x^3+x^{6 r}+x^{4 r}
(1+2 \sqrt{x}+5 x)+x^{2 r} (1+2 \sqrt{x}+2 x+2 x^{3/2}) +\ldots  $  \\
\hline
\caption{Superconformal index for $U(N)$ gauge theories with a symmetric tensor flavor and
a superpotential.\label{tb-index-USym}}
\end{longtable}

The chiral ring structure of $U(N)$ with a symmetric tensor is
almost the same as that of $U(N)$ with an antisymmetric tensor except
that singlets $P_l$ and $\tilde{P}_l$ are symmetric in  flavor
indices rather than antisymmetric. The generators of the chiral ring
make the following contributions to the index:
$\textrm{tr}(X\tilde{X})^i \sim 1\cdot x^{\frac{2i}{n+1}}$,
$i=1,\cdots,n$, $(M_j)^a_b \sim N_f^2\cdot x^{2r+\frac{2j}{n+1}}$,
$j=0,\cdots,n$, $(P_l)^{ab}, (\tilde{P}_l)_{ab}\sim
\frac{N_f(N_f+1)}{2}\cdot x^{2r+\frac{2l+1}{n+1}}$,
$l=0,\cdots,n-1$.

\subsection{$Sp(2N)$ with an adjoint}

We propose the following duality:

\begin{itemize}

 \item Electric theory:  $Sp(2N_c)_k$ gauge theory with $2N_f$ fundamental chiral multiplets $Q^a$, an adjoint chiral multiplet $X$, and a superpotential $W_e=\Tr\, X^{2(n+1)}$

\item Magnetic theory: $Sp(2\big((2n+1)(N_f+k)-N_c-1\big))_{-k}$ gauge theory with $2N_f$ fundamental chiral multiplets $q_a$, singlets $M_{j}^{ab}=Q^a J X^{j} Q^b$, $j=0,\ldots,2n$ which are symmetric (resp. antisymmetric) in their flavor indices for odd (resp. even) $j$, an adjoint chiral multiplet $Y$, and a superpotential
$$W_m=\Tr\, Y^{2(n+1)}+\sum_{j=0}^{2n} M^{ab} q_a JY^{2n-j} q_b$$

\end{itemize}

Chiral ring generators map as follows:
\begin{eqnarray}
Q^a J X^{j} Q^b & \ra & M_{j}^{ab},\quad j=0,\ldots, 2n\\
\Tr\, X^{2j} & \ra & \Tr\, Y^{2j},\quad j=1,\ldots, n.
\end{eqnarray}
R-charge assignment are $R(Q)=r, R(q)=\frac{1}{n+1}-r, R(M_j)=\frac{j}{n+1}+2r, R(X)=R(Y)=\frac{1}{n+1}$. Note that since $XJ$ is symmetric, $\Tr\, X^{2j+1}$ vanishes for any integer $j$. Note also that for $n=0$ this duality reduces to the symplectic 3d Seiberg duality.

We have computed the superconformal index for several dual pairs, the results are displayed in Table \ref{tb-index-SpAd}. In all cases the duality is confirmed.
\begin{longtable}{|c|c|c|p{8cm}|}
\hline
$(N_f,k,N_c)$ & Electric    &   Magnetic     &   \\
$n=1, N_f+k=1$ & $Sp(2N_c)$    & $Sp(2(2-N_c))$    & ~ Index  \\
\hline
(0,1,1)       &   $Sp(2)$  &  $Sp(2)$         &  $  1  $  \\
\hline
\hline
$n=1, N_f+k=2$ & $Sp(2N_c)$    & $Sp(2(5-N_c))$ & \\
\hline
(0,2,2)       &   $Sp(4)$  &  $Sp(6)$         &  $  1 + x + x^3 +\ldots $  \\
\hline
(1,1,2)       &   $Sp(4)$  &  $Sp(6)$         &  $1+x-4 x^2-8 x^{5/2}-9 x^3-x^{3-2 r}+x^{6 r}+x^{4 r} (1+3 \sqrt{x}+8 x)+x^{2 r} (1+3 \sqrt{x}+2 x+3 x^{3/2}-3 x^2) +\ldots  $  \\
\hline
(1,1,3)       &   $Sp(6)$  &  $Sp(4)$         &  $  1 + x - 4 x^2 + x^{4 r} + x^{2 r} (1 + 3 \sqrt{x} + 2 x) +\ldots $  \\
\hline
\hline
$n=2, N_f+k=1$ & $Sp(2N_c)$    & $Sp(2(4-N_c))$ &\\
\hline
(0,1,1)       &   $Sp(2)$  &  $Sp(6)$         &  $ 1 + x^{2/3}  +\ldots$   \\
\hline
(0,1,2)       &   $Sp(4)$  &  $Sp(4)$         &  $  1 + x^{2/3} + x^{4/3} + x^{8/3} +\ldots  $  \\
\hline
\caption{Superconformal index for $Sp(2N)$ gauge theories with an adjoint and a superpotential.\label{tb-index-SpAd}}
\end{longtable}

Chiral ring generators make identifiable contributions to the index: $\Tr\, X^{2i} \sim x^{\frac{2i}{n+1}}$, $M_{2j}^{ab} \sim N_f(2N_f-1)\cdot x^{2r+\frac{2j}{n+1}}$, $M_{2j+1}^{ab}\sim N_f(2N_f+1)\cdot x^{2r+\frac{2j+1}{n+1}}$. Note that the theory with $(n,N_f,k,N_c)=(1,0,1,1)$ has a trivial index $I=1$. Technically, this happens because the contributions of all BPS states are exactly canceled by monopole operators.  In fact, the whole theory is almost trivial in the IR limit (reduces to a TQFT) as a consequence of the Jafferis-Yin duality \cite{JafferisYin11}. Indeed, the Jafferis-Yin duality says that modulo a topological sector $Sp(2)_1$ with an adjoint  $X$ is equivalent to a free theory whose only chiral superfield $u$ is dual to $\Tr\, X^2$. Under this duality the superpotential deformation $W_e=\Tr\, X^4$ is mapped  to the mass term for $u$. Thus the theory with the superpotential is IR trivial modulo a topological sector (which is described by a $U(1)_2$ Chern-Simons theory).

\subsection{$O(N)$ with a symmetric traceless tensor}

We propose the following duality:

\begin{itemize}

 \item Electric theory:  $O(N_c)_k$  gauge theory with $N_f$ fundamental chiral multiplets $Q^a$, a symmetric traceless chiral multiplet $X$, and a superpotential $W_e=\Tr\, X^{n+1}$

\item Magnetic theory:  $O(n(N_f+k+2)-N_c)$ gauge theory with $N_f$ fundamental chiral multiplets $q_a$, $n N_f(N_f+1)/2$ singlets $M_{j}^{ab}$, $j=1,\ldots,n$,  a symmetric traceless chiral multiplet $Y$, and a superpotential
$$
W_m=\Tr\, Y^{n+1}+\sum_{j=1}^{n} M^{ab} q_a Y^{n-j} q_b.
$$

\end{itemize}

Chiral ring generators map as follows:
\begin{eqnarray}
Q^a X^{j-1} Q^b & \ra & M_{j}^{ab},\quad j=1,\ldots,n,\\
\Tr\, X^j & \ra & \Tr\, Y^j,\quad j=2,\ldots,n.
\end{eqnarray}
R-charge assignments are $R(Q)=r, R(q)=\frac{2}{n+1}-r, R(M_j)=\frac{2(j-1)}{n+1}+2r, R(X)=R(Y)=\frac{2}{n+1}$. Note for $n=1$ this duality reduces to the 3d orthogonal Seiberg duality of \cite{Kapustin11}.

We have computed the superconformal index for several dual pairs, the results are displayed in Table \ref{tb-index-OSym}. In all cases the duality is confirmed.
\begin{longtable}{|c|c|c|p{8cm}|}
\hline
$(N_f,k,N_c)$& Electric    &   Magnetic     &  \\
$n=2, N_f+k=1$ & $ O(N_c)$    & $O(6-N_c)$    & ~ Index \\
\hline
(0,1,1)       &   $O(1)$  &  $O(5)$         &  $  1  $  \\
\hline
(0,1,2)       &   $O(2)$  &  $O(4)$         &  $ 1 + x^{4/3} - x^2 + x^{8/3} - x^4 +\ldots  $  \\
\hline
(0,1,3)       &   $O(3)$  &  $O(3)$         &  $ 1+x^{4/3}+x^{10/3}-x^4 +\ldots $  \\
\hline
\hline
$n=2, N_f+k=2$ & $ O(N_c)$    & $O(8-N_c)$    &  \\
\hline
(0,2,3)       &   $O(3)$  &  $O(5)$         &  $ 1+x^{4/3}+2 x^{10/3}-x^4 +\ldots  $  \\
\hline
(1,1,3)       &   $O(3)$  &  $O(5)$         &  $1+x^{4/3}-x^2+x^{4 r}+x^{2 r} (1+x^{2/3}+x^{4/3}) +\ldots  $  \\
\hline
(0,2,4)       &   $O(4)$  &  $O(4)$         &  $ 1 + x^{4/3} + x^{8/3} + x^{10/3} +\ldots  $  \\
\hline
(1,1,4)       &   $O(4)$  &  $O(4)$         &  $ 1 + x^{4/3} - x^2 + x^{8/3} - x^{10/3} - 3 x^4 + x^{6 r} +  x^{4 r} (1 + x^{2/3} + 2 x^{4/3}) +  x^{2 r} (1 + x^{2/3} + x^{4/3} + x^2) + x^{-2 r} (-x^4 - x^{14/3}) +\ldots  $  \\
\hline
(1,1,5)       &   $O(5)$  &  $O(3)$         &  $ 1+x^{4/3}-x^2-x^{\frac{10}{3}-2 r}+x^{4 r}+x^{2 r} (1+x^{2/3}+x^{4/3}) +\ldots  $  \\
\hline
\hline
$n=2, N_f+k=3$ & $ O(N_c)$    & $O(10-N_c)$    &  \\
\hline
(0,3,5)       &   $O(5)$  &  $O(5)$         &  $ 1+x^{4/3}+x^{8/3} +\ldots  $  \\
\hline
(1,2,5)       &   $O(5)$  &  $O(5)$         &  $ 1+x^{4/3}-x^2+x^{4 r}+x^{2 r} (1+x^{2/3}+x^{4/3}) +\ldots  $  \\
\hline
(2,1,5)       &   $O(5)$  &  $O(5)$         &  $ 1+x^{4/3}-4 x^2-3 x^{8/3}+6 x^{4 r}+x^{2 r} (3+3 x^{2/3}+3 x^{4/3}) +\ldots  $  \\
\hline
\hline
$n=3, N_f+k=1$ & $ O(N_c)$    & $O(9-N_c)$    &  \\
\hline
(0,1,3)       &   $O(3)$  &  $O(6)$         &  $  1 + x + x^{3/2} + 2 x^3 +\ldots $  \\
\hline
(0,1,4)       &   $O(4)$  &  $O(5)$         &  $  1 + x + x^{3/2} + x^2 + x^3 +\ldots $  \\
\hline
\caption{Superconformal index for $O(N)$ gauge theories with a symmetric traceless tensor and a superpotential.\label{tb-index-OSym}}
\end{longtable}

Chiral ring generators make identifiable contributions to the superconformal index: $\Tr\, X^i\sim 1\cdot x^{\frac{2i}{n+1}}$, $i=2,\cdots,n$ and $(M_j)^{ab}\sim \frac{N_f(N_f+1)}{2}\cdot x^{2r +\frac{2(j-1)}{n+1}}$, $j=1,\cdots,n$. There are many cases where the chiral ring is truncated nonperturbatively. For example, if the electric gauge group is $O(1)$, the traceless symmetric $X$ vanishes, so the operators $\Tr\, Y^j$ on the magnetic side must be truncated. The extreme case of this is when $N_f=0$, when the whole theory must be IR trivial. On the magnetic side, the contributions to the index coming from the classical generators of the chiral ring are canceled by monopole operators.

\section{Concluding remarks}

We have proposed and tested seven classes of dualities involving $\cN=2$ Chern-Simons gauge theories with tensor and fundamental matter and a superpotential. In the first three cases ($U(N)$ with an adjoint, $SO(2N)$ with an adjoint, and $Sp(2N)$ with an antisymmetric tensor) we found that for a special value of the Chern-Simons coupling the theory with $N_f=0$ and no superpotential is dual to a free theory. In the remaining four cases ($U(N)$ with an (anti)symmetric tensor flavor, $Sp(2N)$ with an adjoint and $O(N)$ with a symmetric traceless tensor) one can argue that such a description does not exist, for any value of the Chern-Simons coupling. To see this, note that far along the moduli space each of these theories reduces to a free CFT describing the moduli and a Chern-Simons TQFT whose gauge group is the unbroken part of the UV gauge group. For example, $Sp(2N)_k$ theory with an adjoint reduces to a theory of $N$ free fields times $N$ copies of $U(1)$ gauge theory at level $2k$. Since the TQFT part is nontrivial for all $k$ and $N$, the $Sp(2N)_k$ theory  with an adjoint cannot be dual to a free theory of moduli for any $N$ or $k$.

In a similar spirit, consider the $O(N)$ theory with a traceless symmetric tensor and no superpotential. Far along the moduli space the gauge group is broken down to a discrete subgroup isomorphic to $O(1)^{N}$ which is again nontrivial. Thus at best one could hope that the full theory is a product of a free CFT describing the moduli and a topological gauge theory with a discrete gauge group. However, even this is not possible. Indeed, if this were true, we would expect that supersymmetry is spontaneously broken whenever the theory is perturbed by a superpotential $\Tr\, X^j$ with $j<N+1$. But the rank of the magnetic gauge group indicates that this happens only for $j<1+N/(k+2)$, which is strictly smaller.

Similar arguments can be used to rule out a free dual for $U(N)$ theory with a symmetric or antisymmetric tensor flavor (excepting of course the case $G=U(1)$ with an antisymmetric tensor and $N_f=0$).

In this paper we only considered dualities for theories with nonzero Chern-Simons couplings. It would also be interesting to find dualities for similar theories with no Chern-Simons couplings, along the lines of \cite{Aharony97,HPP11}.

\vskip 0.5cm  \hspace*{-0.8cm} {\bf\large Acknowledgements} \vskip
0.2cm

\hspace*{-0.75cm}  A. K. is supported in part by the DOE grant DE-FG02-92ER40701. J.P. is supported by the KOSEF Grant
R01-2008-000-20370-0, the National Research Foundation of Korea
(NRF) Grants No. 2009-0085995 and 2005-0049409 through the Center
for Quantum Spacetime (CQUeST) of Sogang University. J. P. also
appreciates APCTP for its stimulating environment for research.
A. K. and J. P. acknowledge Simons Summer Workshop on Mathematics and
Physics 2011 for hospitality while the current work was initiated.

\end{document}